\documentclass[10pt,fleqn,onecolumn,aps,pre,groupedaddress,a4paper,floatfix]{revtex4}

\usepackage{amsmath}
\usepackage{graphicx,psfrag}
\usepackage{multirow}
\usepackage{graphics}
\usepackage{amsfonts}
\usepackage{amssymb}
\usepackage{subfigure}
\usepackage{bbm}
\usepackage{natbib}

\begin{document}
 \title{Non-Markovian decoherence and disentanglement scenarios \\in quantum Brownian motion}
 \author{Christian H\"orhammer}
 \email{christian.hoerhammer@uni-bayreuth.de}
 \affiliation{Theoretische Physik I, Universit\"at Bayreuth, D-95440 Bayreuth, Germany}
 \author{Helmut B\"uttner}
 \affiliation{Theoretische Physik I, Universit\"at Bayreuth, D-95440 Bayreuth, Germany}
 \date{\today}
 \begin{abstract}
We study the non-Markovian decoherence and disentanglement dynamics of dissipative quantum systems with special emphasis on non-Gaussian continuous variable systems. The dynamics are described by the Hu-Paz-Zhang master equation of quantum Brownian motion. The time evolution of the decoherence function of a single-mode superposition is compared to the concurrence of a two-mode entangled state. It is verified that moderate non-Markovian influences slow down the decay of interference fringes and quantum correlations, while strong non-Markovian effects resulting from an out-of-resonance bath can even accelerate the loss of coherence, compared to predictions of Markovian approximations. Qualitatively different scenarios including exponential, Gaussian or algebraic decay of the decoherence function are analyzed. It is shown that partial revivals of coherence can occur in case of non-Lindblad-type dynamics.
\end{abstract}
\maketitle
\section{Introduction}
During the last decade quantum information and computation has been extended from discrete systems to quantum systems with continuous variables such as position and momentum or the amplitudes of electromagnetic field modes. This quantum information theory of continuous variable systems has received much attention in the past few years \cite{braunstein2, braunstein, cerf} and has found various applications in quantum cryptography and quantum teleportation \cite{furusawa, kimble}. Great advances have been made in characterizing the entanglement properties of two-mode Gaussian states by determining the necessary and sufficient criteria for their separability \cite{simon, duan} and by developing quantitative entanglement measures \cite{vidal, wolf}. Non-Gaussian continuous variable states are more difficult to treat theoretically and to be controlled experimentally. Therefore the got much less attention in recent years. However, especially the class of entangled coherent states offers the possibility to apply concepts such as concurrence, first developed for discrete systems and to study the non-Markovian disentanglement dynamics of these states.\\

Due to the unavoidable interaction with the environment, any pure quantum state used in some quantum information process evolves into a mixed state. Thus, a realistic analysis of continuous variable quantum channels must take decoherence and dissipation into account. Decoherence describes the environment-induced suppression of the quantum mechanical coherence properties and interference ability. This concept is strongly related to the measurement problem \cite{zeh1, zeh2, zurek0, schlosshauer1} and the transition from the quantum to the classical regime \cite{zurek3, jooszeh, bolivar, halliwell, Zeh}. The time scales on which these processes take place and strategies to reduce these effects are a major topic of research. Thereby, the theoretical results strongly depend on the underlying dissipative dynamics and the performed approximations.\\

Within the theory of open quantum systems \cite{breuer, dittrich} the dissipative dynamics are mainly described by master equations of the reduced density matrix. Initial quantum superpositions are destroyed and quantum correlations are lost during characteristic decoherence and separability time scales. The Markovian time evolution of quantum correlations of entangled two-mode continuous variable states has been examined in single-reservoir \cite{prauzner, braun} and two-reservoir models \cite{duan, halliwell, olivares2, serafini2}, representing noisy correlated or uncorrelated Markovian quantum channels. Quantum correlations are found to be better preserved in a common reservoir. Additionally the coupling to the same bath variables might generate new quantum correlations between the parts of the subsystem \cite{hoebue, plenhuelgo, benatti1, benatti2, rajagopal2}. The underlying Born-Markov approximation assumes weak coupling between the system and the environment to justify a perturbative treatment and neglects short-time correlations between the system and the reservoir. This approach has been widely and successfully employed in the field of quantum optics \cite{walls} where the characteristic time scales of the environmental correlations is much shorter compared to the internal system dynamics. Challenged by new experimental evidence a growing interest in non-Markovian descriptions can be observed. Very recently some phenomenological \cite{ban2, mcaneney} and microscopic models \cite{manisc6, an, liu, an2} of non-Markovian quantum channels have been proposed. Using the analogy between the Hilbert space of quantized electromagnetic fields and the Hilbert space of quantum harmonic oscillators, the Caldeira Leggett model of quantum Brownian motion \cite{caldeira3, ford3, paz} can be extended to describe the entanglement dynamics of two-mode squeezed states or two-mode entangled coherent states as examples of Gaussian and non-Gaussian states respectively. \\

Non-Markovian effects on decoherence and disentanglement dynamics become important if the decoherence time scale and the bath correlation time scales are comparable. This might be the case for macroscopic superposition since the decoherence time is reduced for increased separation in phase space. The explicit course of the decoherence process depends on the relation of the characteristic time scales of system $\tau_s$ and environment $\tau_b$ on the one hand and the decoherence time scale $\tau_d$ on the other hand. Depending on the relation of the characteristic time scales (including the relaxation time scale $\tau_{\gamma}$ four different regimes can be distinguished:
\begin{eqnarray}
\tau_b\ll&\tau_s&\ll\tau_{\gamma}\qquad\qquad\mbox{(Born-Markovian regime)},\\
\tau_b\approx&\tau_s&\ll\tau_{\gamma}\qquad\qquad\mbox{(Non-Markovian regime)},\\
\tau_b\ll&\tau_s&\approx\tau_{\gamma}\qquad\qquad\mbox{ ({\sl
Strong-coupling} regime)},\\
\tau_s<&\tau_b&\ll\tau_{\gamma}\qquad\qquad\mbox{({\sl
Out-of-resonance} regime)}.
\end{eqnarray}
Since the decoherence time of superposed coherent states depends on the phase space separation, the initial preparation mainly determines the ratio of the decoherence time to the other characteristic time scales. We will focus on the non-Markovian regime where different decoherence scenarios exist -- for both, the dynamics of single-mode superpositions as well as the disentanglement of two-mode entangled coherent states.\\

The paper is organized as follows. In section II, we briefly review the properties of entangled coherent states a a special class of non-Gaussian continuous variable states and introduce relevant decoherence and entanglement measures. In section III, we describe the Hu-Paz-Zhang (HPZ) master equation of quantum Brownian motion which is the basis for studying non-Markvoian effects. We resume the extended, two-mode version of the Caldeira-Leggett model for single and two-reservoir models. In section IV we present and discuss the numerical results of the decoherence and entanglement dynamics of a single-mode superposed coherent state compared to a two-mode entangled coherent state. Different scenarios are analyzed, including exponential, Gaussian and algebraic decay as well as revivals of decoherence. It is shown that the behavior of the concurrence of an entangled two-mode state is equivalent to the behavior of the decoherence function of a single-mode superposition state. Finally, a brief summary is given in section V.
\section{Non-Gaussian continuous variable states}
Great advances have been made in characterizing the entanglement properties of two-mode Gaussian states by determining the necessary and sufficient criteria for their separability \cite{simon, duan} and by developing quantitative entanglement measures \cite{vidal, wolf}. Non-Gaussian continuous variable states are more difficult to treat theoretically and to be controlled experimentally. Therefore the got much less attention in recent years. However, especially the class of entangled coherent states offers the possibility to apply concepts such as concurrence, first developed for discrete systems and to study the non-Markovian disentanglement dynamics of these states.
\subsection{Entangled coherent states (ECS)}
In the following we consider entangled coherent states \cite{sanders1, sanders2} as a special class of non-Gaussian continuous variable states. An example of a multi-mode entangled coherent state is given by
\begin{equation}
|\alpha,\theta,N\rangle=\frac{|\alpha_+\rangle_1\otimes|\alpha_+\rangle_2\otimes...\otimes|\alpha_+\rangle_N+
e^{i\theta}|\alpha_-\rangle_1\otimes|\alpha_-\rangle_2\otimes...\otimes|\alpha_-\rangle_N}{\sqrt{2+2e^{-2N|\alpha_0|^2}\cos\theta}}.
\end{equation}
which is a superposition of coherent states $|\alpha_{\pm}\rangle_i=\hat
D(\pm \alpha_0)|0\rangle_i$ with displacement parameter $|\alpha_0|^2=\frac{q_0^2}{4\sigma_0^2}+\frac{\sigma_0^2p_0^2}{\hbar^2}$ and $\sigma_0$ as width of a minimum uncertainty wave packet displaced at $q_0$, $p_0$ (mean values) in phase space. In particular, we will focus on the two-mode version
\begin{equation}\label{phialphacat}
|\Phi_{\pm}\rangle:=|\alpha,\theta=\{0,\pi\},N=2\rangle=\frac{\left(|\alpha_+\rangle_1|\alpha_+\rangle_2\pm |\alpha_-\rangle_1|\alpha_-\rangle_2\right)}{\sqrt{2(1\pm e^{-4|\alpha_0|^2})}}
\end{equation}
which is also known as quasi-Bell-state \cite{hirota3, hirota4, an} due to its similarity to the well-known Bell states as maximally entangled two-qubit states. In addition to a violation of the Bell-inequalities \cite{sanders3} these states show further nonclassical properties, such as sub-Poissonian statistics,  squeezing and a violation of the Cauchy-Schwarz inequaltiy \cite{chai}. Contrary to entangled single Fock-states \cite{tanwalls} ECS are superpositions of non-orthogonal states \cite{wang} with some remarkable properties \cite{hirota1, hirota2, lixu}.\\
If just a single mode is considered, we get the well known cat-state superposition of two coherent states
\begin{equation}\label{cohcat}
|\psi_{\alpha}\rangle=\frac{|\alpha_+\rangle+e^{i\theta}|\alpha_{-}\rangle}{\sqrt{2+2
e^{-2|\alpha_0|^2}\cos\theta}}\end{equation}
(which is of course not entangled). Its density matrix can be separated into two parts,
\begin{equation}
\rho_{\alpha}(0)=\underbrace{\frac{|\alpha_+\rangle\langle
\alpha_+|+|\alpha_-\rangle\langle
\alpha_-|}{2(1+e^{-2|\alpha_0|^2})}}_{\rho_{cl}(0)}+\underbrace{\frac{|\alpha_-\rangle\langle
\alpha_+|+|\alpha_+\rangle\langle
\alpha_-|}{2(1+e^{-2|\alpha_0|^2})}}_{\rho_I(0)},
\end{equation}
a classical part $\rho_{cl}$ and a part $\rho_I$ describing the quantum mechanical ability to interfere (here $\theta=0$). The corresponding phase space representation in form of the Wigner function
\begin{equation}
\label{Wcat}
W_{\pm q_0}(q,p,0)=\frac{e^{-q^2/2\sigma_0^2-2p^2\sigma_0^2/\hbar^2}}{\pi\hbar\left(1+e^{q_0^2/2\sigma_0^2}\right)}
\left(e^{q_0^2/2\sigma_0^2}\cos\frac{2q_0p}{\hbar}+\cosh\frac{q_0q}{\sigma_0^2}\right),
\end{equation}
(for $|\alpha_0|^2=\frac{q_0^2}{4\sigma_0^2}$) has two characteristic peaks at $q=\pm q_0$ and an oscillating pattern in between with partially negative values of the Wigner function. It is one of the most common examples for studying decoherence for continuous variable states. In this paper we want to compare the decoherence properties of these state within a non-Markovian description to the disentanglement process of an two-mode entangled coherent state.
\subsection{Decoherence and Disentanglement measures}
\subsubsection{Decoherence of a single-mode superposition state}\quad\\
There are different measures of decoherence in phase space, in position and momentum space or Fock space representation. Decoherence time scales based on different measures can deviate from each other. In our view, phase space measures give the better picture of the underlying non-Markovian process compared to a treatment in position space alone (i.a. the so-called attenuation factor or the fringe visibility function). Decoherence in phase space can be described by the time evolution of the purity $\mu(t)=\mbox{Tr}[\rho^2(t)]$ or the so called quantumness $\Xi(t)=\int dqdp |W(q,p,t)|-1$ which measures the phase space volume of the negative part of the Wigner function. While the first mixes decoherence and relaxation phenomena and the latter is difficult to calculate, the norm $\mu_I(t)=\mbox{Tr}[\rho_I^2(t)]$ of the interference part seems to be an adequate measure for decoherence \cite{haake1}. It can be calculated from the oscillating part $W_I(q,p,t)$ of the Wigner function by
\begin{equation}
\mu_I(t)=(2\pi\hbar)^2\int_{-\infty}^{\infty}dq \int_{-\infty}^{\infty}dp\, W_I^2(q,p,t).
\end{equation}
The non-Markovian time evolution of the Wigner function is governed by a quantum Fokker-Planck equation given in the third section.
\subsubsection{Disentanglement of a two-mode entangled coherent state}\quad\\
To study the entanglement properties of a non-Gaussian continuous variable state we use the concept of concurrence \cite{wootters1, wootters2}. For the density matrix $\rho_{12}$ of a pair of qubits the concurrence is defined as
$\mathcal{C}_{12}=\mbox{max}\left\{\lambda_1-\lambda_2-\lambda_3-\lambda_4,0\right\}$,
where $\lambda_1\geq\lambda_2\geq\lambda_3\geq\lambda_4$ are the square roots of the eigenvalues of the spin-flipped state
$\varrho_{12}\equiv\rho_{12}\left(\sigma_y\otimes\sigma_y\right)\rho_{12}^*\left(\sigma_y\otimes\sigma_y\right)$ with Pauli matrix $\sigma_y$. The concurrence varies from $\mathcal{C}_{12}=0$ for disentangled qubits $1$ and $2$
to $\mathcal{C}_{12}=1$ for maximally entangled states. To determine the pairwise entanglement of a $N$-mode system in analogy to the discrete $N$-qubit system the reduced density matrix $\rho_{12}=\mbox{Tr}_{2,4,..N}\left\{|\alpha,\theta,N\rangle\langle\alpha,\theta,N|\right\}$ of two modes $1$ and $2$ has to be considered (all reduced density matrices $\rho_{kl}$ of any two modes $k$ and $l$ are identical). The trick is to represent the density matrix
{\mathindent0.5cm
\begin{eqnarray}
\rho_{12}&=&\frac{1}{2(1+e^{-4N|\alpha|^2})}\left\{|\alpha_+\rangle|\alpha_-\rangle\langle\alpha_+|\langle\alpha_-|\,+\,
|\alpha_-\rangle|\alpha_+\rangle\langle\alpha_-|\langle\alpha_+|+\right.\nonumber\\
&&\left.\,e^{i\theta-2(N-2)|\alpha|^2}|\alpha_-\rangle|\alpha_+\rangle\langle\alpha_+|\langle\alpha_-|\,+\,
e^{-i\theta-2(N-2)|\alpha|^2}|\alpha_+\rangle|\alpha_-\rangle\langle\alpha_-|\langle\alpha_+|\right\}
\end{eqnarray}}
in an orthogonal basis $\{|{\bf0}\rangle,|{\bf1}\rangle\}$ with
$
|{\bf0}\rangle\equiv|\alpha_+\rangle$ and $ |{\bf1}\rangle\equiv\frac{|\alpha_-\rangle-e^{-2|\alpha|^2}|\alpha_+\rangle}{\sqrt{1-e^{-4|\alpha|^2}}}
$,
\begin{equation}
\rho_{12}=N_0^2\left(\begin{array}{cccc}2p^2(1+q\cos\theta)&pM(1+qe^{i\theta})&pM(1+qe^{-i\theta})&0\\
pM(1+qe^{-i\theta})&M^2&M^2qe^{-i\theta}&0\\pM(1+qe^{i\theta})&M^2qe^{i\theta}&M^2&0\\0&0&0&0\end{array}\right)
\end{equation}
with parameters $p=e^{-2|\alpha|^2}$, $N_0=(2+2p^N\cos\theta)^{-1}$,  $q=p^{N-2}$ and $M=\sqrt{1-p^2}$ \cite{sanders2}. By determining the four eigenvalues $\lambda_1=N_0^2M^2(1+q)$, $\lambda_2=N_0^2M^2(1-q)$, and $\lambda_3=\lambda_4=0$ the concurrence of the reduced state $\rho_{12}$ can be calculated to
\begin{equation}\label{concurN}
\mathcal{C}_{12}=\frac{M^2q}{1+p^N\cos\theta}=\frac{e^{-2N|\alpha|^2}(1-e^{4|\alpha|^2})}{1+e^{-2N|\alpha|^2}\cos\theta}.
\end{equation}
In case of a two-mode state ($N=2$) we have pure state entanglement. The entanglement dynamics are then described by the time evolution of the parameters $p(t)$, $q(t)$ and $M(t)$. A specific model will be introduced in the next section.

\section{Non-Markovian dynamics of continuous variable states}
\subsection{HPZ master equation of quantum Brownian motion}
Non-Markovian effects are discussed here on the basis of the Caldeira-Leggett model of quantum Brownian motion \cite{caldeira1,
caldeira2, caldeira3} often referred to as independent-oscillator-model \cite{ford4, ford3}. It is a system plus reservoir model where the total Hamiltonian $H=H_s+H_b+H_{\rm int}$ consists of three parts, with $H_s$ as Hamiltonian of the subsystem which interacts via the Hamiltonian $H_{\rm int}$
with a bath that is described by a collection of a large number of harmonic oscillators $H_b=\sum_i\hbar\omega_i(b^{\dagger}b+1)$.
In detail the Hamiltonian of the Caldeira Leggett model is given by
\begin{equation}
\label{Hcl}H=\frac{p^2}{2m}+V(q)+\sum_{i=1}^N\left[\frac{p_i^2}{2m_i}+\frac{m_i\omega_i^2}{2}
\left(x_i-\frac{c_iq}{m_i\omega_i^2}\right)^2\right],
\end{equation}
where $q$ and $p$ are the Heisenberg-operators of coordinate and momenta of the Brownian oscillator moving in an harmonic potential $V(q)=\frac{1}{2}m\omega_0^2q^2$ and coupled to a bath of $N$ independent harmonic oscillators with variables $x_i$, $p_i$ and frequencies $\omega_i$. The bath is characterized by its spectral density
\begin{equation}\label{Jdrude}
J(\omega)=\pi\sum_{i=1}^N\frac{c_i^2}{2m\omega_i}\delta(\omega-\omega_i)=\frac{\gamma\omega\Gamma^2}{\omega^2+\Gamma^2}.\end{equation}
The interaction is bilinear in the coordinates $q$ and
$x_i$ of the subsystem and the bath respectively with coupling parameters $c_i$. The self-interaction term
(proportional to $q^2$) in the Hamiltonian
\begin{equation}\label{Hint}
H_{\rm
int}=\sum_i\left[-c_ix_iq+\frac{c_i^2}{2m_i\omega_i^2}q^2\right]
\end{equation}
renormalizes the oscillator potential to ensure that the observable frequency is close to bare oscillator frequency $\omega_0$. From influence functional path integral techniques Hu, Paz, Zhang have derived the master equation
{\mathindent0.5cm\begin{equation}\label{hpzeq}
\dot\rho=\frac{1}{ i\hbar
}\left[H_s,\rho\right]+\frac{m\delta\Omega^2(t)}{2i\hbar}[q^2,\rho]+\frac{\gamma_p(t)}{2i\hbar}[q,\{p,\rho\}]+\frac{D_{qp}(t)}{\hbar^2}[q,[p,\rho]]-\frac{D_p(t)}{\hbar^2}[q,[q,\rho]],
\end{equation}}
with $[\,.\,]$ and $\{\,.\,\}$ denoting commutator and anti-commutator respectively. This master equation is valid for arbitrary coupling and temperature. The non-Markovian character is contained in the time-dependent coefficients which read in expansion up to the second order in the system-bath coupling constant \cite{paz}:
{\mathindent0.5cm
\begin{eqnarray}\label{hpzkoeff1}
\gamma_p(t)&=&\frac{2}{\hbar
m\omega_0}\int_0^tdt'L(t')\sin\omega_0t'\thinspace\stackrel{t\gg\Gamma^{-1}}{\longrightarrow}\thinspace\frac{\gamma}{m}\frac{\Gamma^2}{\omega_0^2+\Gamma^2},\\
\label{hpzkoeff2}
\delta\Omega^2(t)&=&\frac{\gamma\Gamma}{m}-\frac{2}{\hbar
m}\int_0^tdt'L(t')\cos\omega_0
t'\thinspace\thinspace\stackrel{t\gg\Gamma^{-1}}{\longrightarrow}\thinspace\frac{\gamma\Gamma}{m}\left(1-\frac{\Gamma^{2}}{\omega_0^2+\Gamma^2}\right) ,\\
\label{hpzkoeff3}
D_{qp}(t)&=&\frac{1}{m\omega_0}\int_0^tdt'K(t')\sin\omega_0t'\thinspace\stackrel{t\gg\Gamma^{-1}}{\longrightarrow}\thinspace m\gamma_q(\infty)\langle q^2\rangle-\frac{\langle p^2\rangle}{m},\\
\label{hpzkoeff4}
D_p(t)&=&\int_0^tdt'K(t')\cos\omega_0t'\thinspace\stackrel{t\gg\Gamma^{-1}}{\longrightarrow}\thinspace
\langle p^2\rangle\gamma_p(\infty),
\end{eqnarray}}
where $L(t)=i\langle [\eta(t),\eta(0)]\rangle$ and $K(t)=\frac{1}{2}\langle\{\eta(t),\eta(0)\}\rangle$ are connected to the spectral density \eqref{Jdrude} by
\begin{eqnarray}
L(t)&=&\frac{\hbar}{\pi}\int_0^{\infty}d\omega\,
J(\omega)\sin\omega t,\\
K(t)&=&\frac{\hbar}{\pi}\int_0^\infty d\omega\,
J(\omega) \coth(\frac{1}{2}\beta\hbar\omega)\cos\omega t.
\end{eqnarray}
$K(t)$ is the correlation function of the quantum noise term $\eta$ resulting from averaging over the initial thermal bath distribution. The exact expressions of the HPZ-coefficients are related to the Green's functions of the corresponding quantum Langevin equations \cite{karrlein, reibold}. The entanglement properties of the joint state of the oscillator and its environment have been studied in ref. \cite{eisert}.
\subsection{Characteristic time scales}
The characteristics of the decoherence process depend on the relation between the different time scales of the systems on the one hand and the decoherence time scale on the other hand. The characteristic time scales of the system -- the internal system dynamics $\tau_s$, the relaxation time scale $\tau_{\gamma}$ and the bath correlation time scale $\tau_b$ -- are determined by the coefficients of the HPZ master equation \eqref{hpzeq} and can be approximated by
{\mathindent1cm\begin{eqnarray}
\tau_s&\approx&(\omega_0^2+\delta\Omega^2)^{-1/2}=\left[\omega_0^2+\frac{\gamma\Gamma}{m}\left(1-\frac{\Gamma^2}{\omega_0^2+\Gamma^2}\right)\right]^{-1/2}\thinspace\sim\quad\omega_0^{-1},\\
\tau_{\gamma}&\approx&\gamma_p^{-1}=\frac{m}{\gamma}\left(1+\frac{\omega_0^2}{\Gamma^2}\right)\qquad\sim\quad
\gamma^{-1},\\
\tau_b&\approx&\mbox{min}\left\{\Gamma^{-1},\beta\hbar\right\}\qquad\sim\quad
\Gamma^{-1}.
\end{eqnarray}}
The decoherence time scale $\tau_d$ for entangled coherent states is mainly governed by the phase space separation in form of the parameter $|\alpha|$.
\subsection{Secular approximation of the HPZ master equation}
Performing a secular approximation of the HPZ master equation \eqref{hpzeq} by averaging over the rapidly oscillating terms of the time-dependent coefficients \eqref{hpzkoeff1}-\eqref{hpzkoeff4} (which is equivalent to a rotating wave approximation after tracing over the environment without neglecting the counter-rotating terms) one gets the following approximated master equation \cite{manisc1, manisc2, manisc3, manisc4}:
\begin{equation}\label{maniseq}
\dot\rho=-i\omega_0[a^{\dagger}a,\rho]+\frac{\tilde\gamma_{\downarrow}(t)}{2}\left[2
a\rho a^{\dagger}-a^{\dagger}a\rho-\rho
a^{\dagger}a\right]+\frac{\tilde\gamma_{\uparrow}(t)}{2} \left[2
a^{\dagger}\rho a-a a^{\dagger}\rho-\rho a a^{\dagger}\right].
\end{equation}
The form of this equation is similar to the quantum optical master equation of the damped harmonic oscillator in Lindblad form, with the only difference that the coefficients $\tilde\gamma_{\downarrow,\uparrow}$ appearing in the master equation are time-dependent. The connection to the HPZ-coefficients \eqref{hpzkoeff1} and
\eqref{hpzkoeff4} is given by
\begin{eqnarray}\label{gammatilde}
\tilde\gamma_{\downarrow}(t)&=&\left(\frac{D_p(t)}{\hbar
m\omega_0}+\frac{\gamma_p(t)}{2}\right)\thinspace\stackrel{t\gg\Gamma^{-1}}{\longrightarrow}\thinspace\frac{\gamma}{m}\frac{\Gamma^2}{\omega_0^2+\Gamma^2}\left(\bar{n}+1\right),\\
\tilde\gamma_{\uparrow}(t)&=&\left(\frac{D_p(t)}{\hbar
m\omega_0}-\frac{\gamma_p(t)}{2}\right)\thinspace\stackrel{t\gg\Gamma^{-1}}{\longrightarrow}\thinspace\frac{\gamma}{m}\frac{\Gamma^2}{\omega_0^2+\Gamma^2}\bar{n}.
\end{eqnarray}
In the limit $t\gg\Gamma^{-1}$ the reach the corresponding Markovian values adjusted by a factor
$\Gamma^2/(\Gamma^2+\omega_0^2)\approx 1$. As long as the coefficients
$\tilde\gamma_{\downarrow,\uparrow}$ are positive for all times the equation \eqref{maniseq} is of Lindblad type. However, not every master equation of Lindblad type form does necessarily fulfill the semigroup property \cite{manisc2}. For certain parameter ranges the coefficients can become negative and the dynamical evolution is of non-Lindblad type \cite{manisc1}.
\subsection{Two-reservoir model}
The dynamics of two identical, not directly interacting modes (with coordinates and momenta $q_j,p_j$, $j=1,2$) in two uncorrelated reservoirs is modeled by the interaction Hamiltonian
\begin{equation}
H_{\rm int}=-q_1\sum\limits_{i=1}^{\infty}c_ix_i^b-q_2\sum\limits_{i=1}^{\infty}c_ix_i^c
\end{equation}
with $\langle x_i^bx_j^c+x_i^cx_j^b\rangle=0\quad\forall\, i,j$. The master equation of the reduced density matrix is then given by the sum of the master equations of two single modes \cite{manisc6}:
\begin{eqnarray}\label{hpzeq2bath}
\dot\rho=\sum_{j=1}^2\left\{\left[\frac{p_j^2}{2i\hbar
m}+\frac{m\gamma_q(t)q_j^2}{2i\hbar
},\rho\right]+\frac{\gamma_p(t)}{2i\hbar}[q_j,\{p_j,\rho\}]+\frac{D_{qp}(t)}{\hbar^2}[q_j,[p_j,\rho]]-\frac{D_p(t)}{\hbar^2}[q_j,[q_j,\rho]]\right\}.
\end{eqnarray}
The time dependent coefficients are given by $\gamma_q(t)=\omega_0^2+\delta\Omega^2(t)-\gamma\Gamma/m$
and eq. \eqref{hpzkoeff1} to \eqref{hpzkoeff4}. The time evolution of a (Gaussian) two-mode squeezed state in two uncorrelated non-Markovian channels has been studied recently in ref. \cite{manisc6} (while we focus on the evolution of non-Gaussian states). The authors derived a non-Markovian separability function which shows oscillations in case of an artificial {\sl out of resonance} bath with $\Gamma\ll\omega_0$. In this two-reservoir model the initial entanglement is always completely lost and both modes are finally uncorrelated (even at zero temperature while $\tau_1\to\infty$). If the secular approximation is applied, the non-Markovian dynamics can be described by the master equation
\begin{equation}\label{meq2bathnonmarkov}
\frac{\partial \rho}{\partial t}=\frac{\tilde \gamma_{\downarrow}(t)}{2}\sum\limits_{j=1}^2\left[
2a_j\rho a_j^{\dagger}-a_j^{\dagger}a_j\rho-\rho
a_j^{\dagger}a_j\right]+\frac{\tilde\gamma_{\uparrow}(t)}{2}\sum\limits_{j=1}^2\left[2
a_j^{\dagger}\rho a_j-a_j a_j^{\dagger}\rho-\rho a_j a_j^{\dagger}\right],
\end{equation}
with time dependent coefficients given in eq. \eqref{gammatilde}. It should be noticed that for $\bar{n}\to 0$ the coefficient  $\tilde\gamma_{\uparrow}(t)$ is different from zero for short times and vanishes for times $t\gg\Gamma^{-1}$.
\subsection{Solution of the model}
In this section we present the solution of the HPZ master equation for the relevant initial states.
\subsubsection{Decoherence function}\quad\\
Given the characteristic function $\chi_I(\eta,\nu,0)$ of an initial preparation, the time-dependent
Wigner function $W_I(q,p,t)$ can be calculated from \cite{oconnell}:
\begin{equation}\label{Lsgwigner}
W_I(q,p,t)=\frac{1}{(2\pi\hbar)^2}\int_{-\infty}^{\infty}d\eta \int_{-\infty}^{\infty}d\nu \,\tilde\chi_I(\eta_t,\nu_t,0)\,\exp\left[i(\eta p+\nu q)/\hbar\right],
\end{equation}
where $\tilde \chi_I$ is identical to the initial characteristic function by substituting
$\eta_t=\dot f(t)\eta+\frac{1}{m}f(t)\nu$, $\nu_t=m\ddot f(t)\nu+\dot f(t)\eta$ and multiplying a Gaussian factor, \begin{eqnarray}
\tilde \chi_I(\eta_t,\nu_t,0)=\chi_I\left(\dot
f(t)\eta+\frac{1}{m}f(t)\nu,m\ddot f(t)\eta+\dot f(t)\nu;0\right)\exp\left[-\frac{1}{2\hbar^2}(K_p(t)\eta^2+2K_{qp}(t)\eta\nu+K_q(t)\nu^2)\right]
\end{eqnarray}
The coefficients $K_{p,q}(t)$ and $K_{qp}(t)$ as well as the correlations $\langle
q^2(t)\rangle$, $\langle p^2(t)\rangle$ and $\langle\{q,p\}(t)\rangle$
are obtained as solution of the corresponding quantum Langevin equation with Greens function $f(t)$.
With $a=\frac{q_0}{2\sigma_0^2}\dot
f(t)$, $b=\frac{q_0}{2m\sigma_0^2}f(t)$ and $\tilde N_0=(1+e^{q_0^2/\sigma_0^2})^{-1}$ the norm $\mbox{Tr}[\rho_I^2]$ of the interference part finally reads
\begin{equation}\label{mudeco}
\mu_I(t)=\mu_0(t)\tilde N_0+\mu_0(t)\tilde N_0\exp\left[\frac{\mu_0^2(t)q_0^2}{\hbar^2m^2\sigma_0^4}\left(m^2\dot
f^2_t\langle q^2(t)\rangle-mf_t\dot
f_t\langle\{q,p\}\rangle+f^2_t\langle
p^2(t)\rangle\right)\right]
\end{equation}
(normalized to $\mu_I(0)=1$). This is our decoherence function.
\subsubsection{Concurrence}\quad\\
The concurrence of a two-mode entangled coherent state under Markovian evolution in a zero temperature environment with $\gamma_{\downarrow}=\gamma$ and $\gamma_{\uparrow}=0$ can be calculated by introducing a time dependent orthogonal basis.
\begin{equation}
|{\bf0}(t)\rangle\equiv|\alpha_+(t)\rangle,\qquad |{\bf1}(t)\rangle\equiv\frac{|\alpha_-(t)\rangle-e^{-2e^{-\gamma t}|\alpha|^2}|\alpha_+(t)\rangle}{\sqrt{1-e^{-4e^{-\gamma t}|\alpha|^2}}}
\end{equation}
with $|\alpha_{\pm}(t)\rangle=|\pm\alpha_0e^{-\gamma t/2}\rangle$. This leads to a time dependent density matrix $\rho_{12}(t)$ with coefficients $p(t)=e^{-2e^{-\gamma t}|\alpha|^2}$, $q(t)=e^{-4(1-e^{-\gamma t})|\alpha|^2}$, and $M(t)=\sqrt{1-p^2(t)}$. Under non-Markovian evolution the constant coefficients $\gamma_{\downarrow}$, $\gamma_{\uparrow}$ have to be replaced by the time dependent coefficients $\tilde\gamma_{\downarrow}(t)$, $\tilde\gamma_{\uparrow}(t)$ which leads to the substitution
\begin{equation}\gamma t\to \Gamma_p(t)=\int_0^tds\gamma_p(s)\quad\mbox{ and }\quad \Delta(t)\to\Delta_p(t)=\frac{2e^{-\Gamma_p(t)}}{\hbar
m\omega_0}\int_0^tds\,e^{\Gamma_p(s)}D_p(s).\end{equation} Thus, we receive the time dependent coefficients of the density matrix  \begin{equation}
p(t)=e^{-2e^{-\Gamma_p(t)}|\alpha|^2},\quad q(t)=e^{-4\Delta_p(t)|\alpha|^2},\quad M(t)=\sqrt{1-p^2(t)},
\end{equation}
From the corresponding eigenvalues $\lambda_i(t)$, $i=1,...,4$ the concurrence can be calculated
\begin{equation}
\mathcal{C}_{12}(t)=N^2M^2(t)q(t)=\frac{1-e^{-4e^{-\Gamma_p(t)}|\alpha|^2}}{1-e^{-4|\alpha|^2}\cos\theta}e^{-4\Delta_p(t)|\alpha|^2},
\end{equation}
and is compared to the time-dependent decoherence function of a single-mode superposition state in section IV.C.
\section{Decoherence and disentanglement scenarios in the non-Markovian regime}
\subsection{Exponential and Gaussian decay}
\begin{figure}[h]
\begin{center}
\psfrag{t}[c]{$2q_0$}\psfrag{x}[c]{$\tau_d$}
\includegraphics[width=8cm,clip]{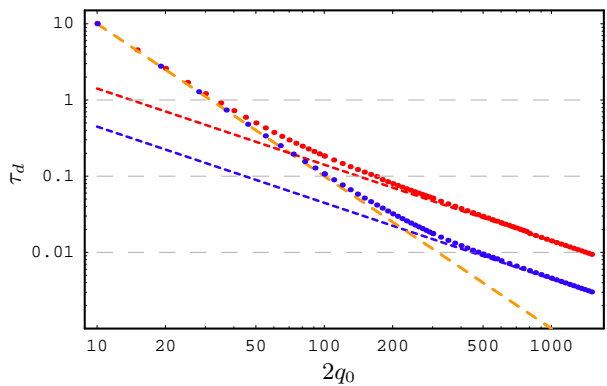}
\caption[] {\label{decoq0}\sl\small Decoherence time scale $\tau_d$ for $\tau_d\ll\tau_s$ at high temperatures in dependence of the separation $|q-q'|=2q_0$. For relatively small values of $q_0$ an thus $\tau_d>\tau_b\approx\Gamma^{-1}$,
 $\tau_d$ decays according to $\hbar^2/(2\gamma kT)q_0^{-2}$ (yellow dashed). If the decoherence time becomes very small $\tau_d\ll\tau_b$, the dependence of $\tau_d$ on $q_0$ changes qualitatively to an inverse proportionality
 $\hbar/(\sqrt{\gamma\Gamma kT})q_0^{-1}$. Here: $\Gamma=10\omega_0$ (read), $\Gamma=100\omega_0$ (blue). Further parameters: $\gamma=10^{-5}m\omega_0$, $kT=50\hbar\omega_0$.}
 \end{center}
\end{figure}

For given system and bath parameters the decoherence time scale varies with the initial phase space separation $|\alpha|$. The parameter $q_0=\frac{1}{2}|q-q'|$ therefore allows to determine the relation between the system dynamics $\tau_s$ and the decoherence timescale $\tau_d$. In the limits $\tau_d\gg\tau_s$ and $\tau_d\ll\tau_s$ it is possible to derive the decoherence time analytically by considering just the leading terms in the HPZ master equation $\eqref{hpzeq}$.\\
For $\tau_s\ll\tau_d\ll\tau_{\gamma}$ the diffusion coefficient $D_p(t)$ dominates the dynamics thus having
\begin{equation}\label{muGR}
\mu_{\tau_d>\tau_s}(t)=\left(\exp\left[-2\frac{q_0^2}{\hbar^2}D_p(\infty)t\right]\right)^2=\exp\left[-\frac{4q_0^2}{\hbar^2}\left(\int_0^{\infty}dt'K(t')\cos\omega_0
t'\right) t\right],
\end{equation}
where the stationary value $D_p(\infty)=\gamma_p(\infty)\langle
p^2(\infty)\rangle$ with $\langle p^2(\infty)\rangle$ is received from the solution of the corresponding Langevin equation
for times $t\approx\Gamma^{-1}\ll\tau_d$. For
$\gamma\ll m\omega_0$ and $\Gamma\gg\omega_0$ we have
$D_p(\infty)=\frac{\gamma\hbar\omega_0}{2}\coth(\frac{1}{2}\beta\hbar\omega_0)$,
thus for $kT\gg\hbar\omega_0$ receiving the decoherence function
\begin{equation}
\mu_{\tau_d>\tau_s}(t)\thinspace\stackrel{\gamma\ll\omega_0}{\approx}\thinspace
\exp\left[-\frac{4\gamma
\omega_0q_0^2}{2\hbar}\coth(\frac{1}{2}\beta\hbar\omega_0)t\right]\thinspace\stackrel{kT\gg\hbar\omega_0}{\longrightarrow}\thinspace
\exp\left[-\frac{4\gamma kTq_0^2}{\hbar^2}t\right].
\end{equation}
Decoherence is dominated by thermal fluctuations and the decoherence function decreases exponentially with characteristic time scale
\begin{equation}
\mu_{\tau_d>\tau_s}(t)\sim\exp[-t/\tau_d]\quad\mbox{with}\quad\tau_d=\frac{\hbar^2}{4\gamma
kT}q_0^{-2}
\end{equation}
and scales inverse quadratically with the displacement $q_0$ \cite{walls3}. \\
In the limit $\tau_d\ll\tau_s\ll\tau_{\gamma}$ decoherence takes place on time scales $t\ll w_0^{-1}$ and is dominated by vacuum fluctuations and system-bath interaction. The decoherence factor can be derived from influence functional path integrals methods  and is given by
\begin{eqnarray}\label{muWW}
\mu_{\tau_d<\tau_s}(t)=\left(\exp\left[-\frac{4q_0^2}{\hbar^2}\int_0^tdt'\int_0^{t'}dt''K(t'-t'')\right]\right)^2=\exp\left[-\frac{8q_0^2}{\pi\hbar}\int_0^{\infty}d\omega
J(\omega) \coth(\frac{1}{2}\beta\hbar\omega)\frac{1-\cos(\omega
t)}{w^2}\right].
\end{eqnarray}
The decay of the decoherence function becomes more and more non-exponential when the parameter $q_0$ is increased and finally becomes Gaussian for $\tau_d\ll\tau_b\ll\tau_s\ll\tau_{\gamma}$ with
\begin{equation}
\mu_{\tau_d<\tau_s}(t)\sim\exp[-(t/\tau_d)^2]\quad\mbox{with}\quad
\tau_d=\frac{\hbar}{2\sqrt{K(0)}}q_0^{-1}\thinspace\stackrel{kT\gg\hbar\omega_0}{\longrightarrow}\thinspace\frac{\hbar}{2\sqrt{\gamma\Gamma
kT}}q_0^{-1},
\end{equation}
where the dissipation-fluctuation theorem $K(t)=kT\gamma(t)=kT\gamma\Gamma e^{-\Gamma t}$ has been applied. In this case the decoherence time scales just linearly with the inverse initial separation $q_0$ and depends explicitly on the bath cut off frequency $\Gamma$. Determining the decoherence time by just considering the diffusion term leads to an underestimation of the decoherence rate. Thus, the decoherence scenarios for mesoscopic ($\tau_s\ll\tau_d\ll\tau_{\gamma}$) and macroscopic separations ($\tau_d\ll\tau_s$) differ from each other \cite{haake1, haake2}. The different power law scaling of $\tau_d$ with respect to the separation $q_0$ is illustrated by figure \ref{decoq0}.
\subsection{Long and short time behavior}
\begin{figure}[h]
\begin{center}
\psfrag{t}[c]{$\omega_0t$}\psfrag{x}[c]{$\mu_I(t)$}
\psfrag{0.05}[c]{}
\subfigure[$kT=10\hbar\omega_0$, $|\alpha_{0}|^2=30$.]{\includegraphics[width=5cm,clip]{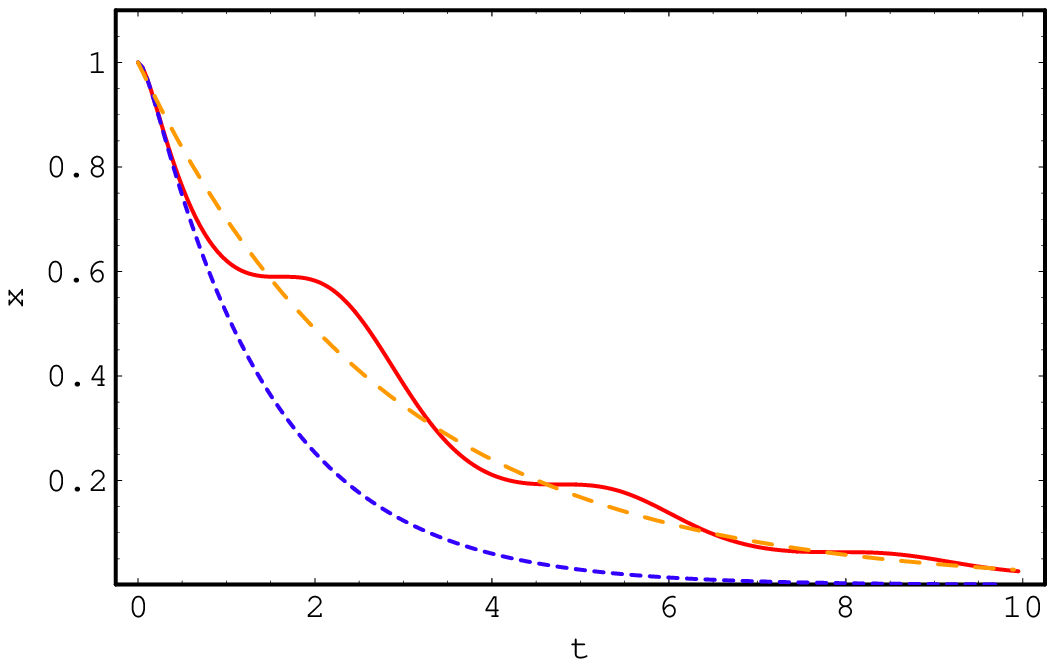}}\qquad
\subfigure[$kT=10^{-3}\hbar\omega_0$, $|\alpha_{0}|^2=100$. ]{\includegraphics[width=5cm,clip]{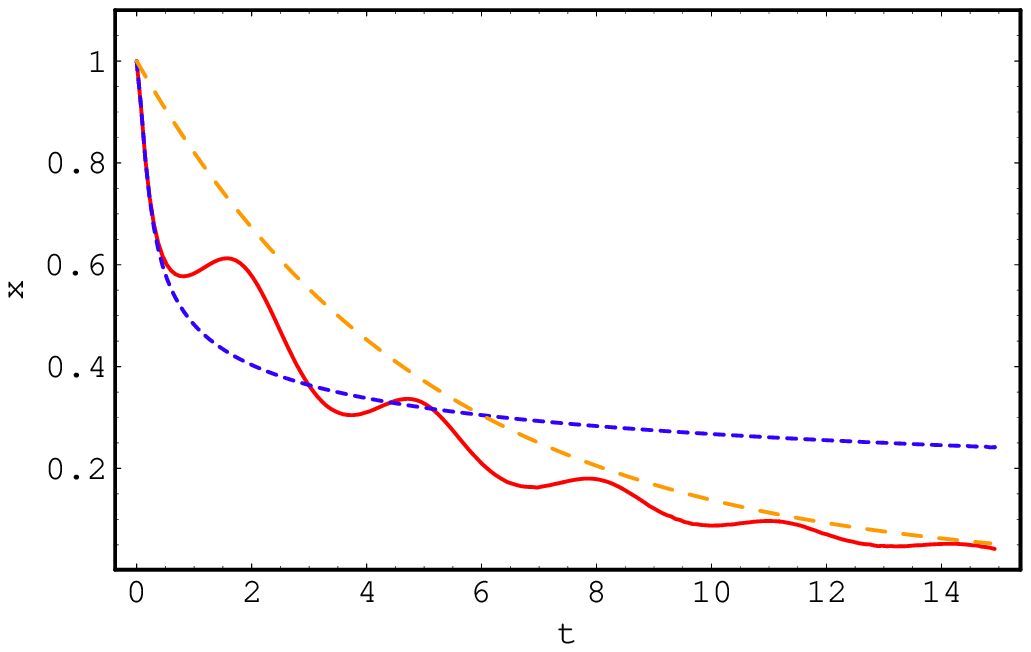}}
 \caption[]{\label{mumediumTh}\sl\small
  Time evolution of the decoherence function \eqref{mudeco} (red line) in comparison to the approximation \eqref{muWW}
 (blue dashed line) and \eqref{muGR} (yellow dashed line) for high temperatures (a) and low temperatures (b). The oscillations result from the rotation and {\sl breathing} of the Wigner function in phase space (with frequency $\omega_0\sim\tau_s^{-1}$) which is related to a periodical change between superpositions in coordinate and momentum space. Further parameters: $\gamma=10^{-5}\omega_0$, $\Gamma=10\omega_0$.}
 \end{center}
\end{figure}

In this section we focus on the decoherence of macroscopic superpositions (large $|\alpha|$) for the standard scenario $\tau_b\ll\tau_s\ll\tau_{\gamma}$ at high and low temperatures. The decoherence time scale $\tau_d$
thus has the same magnitude as the internal system dynamics $\tau_s$. For $kT\gg\hbar\omega_0$ the functions
\eqref{muGR} and \eqref{muWW} become
\begin{equation}
\mu_{\tau_d>\tau_s}(t)=e^{-(4\gamma
kTq_0^2/\hbar^2)t}\qquad\mbox{and}\qquad\mu_{\tau_d<\tau_s}(t)=e^{-(8\gamma
kTq_0^2/\hbar^2)t}.
\end{equation}
These approximated functions decay exponentially and differ just by a factor of two in the exponent. From figure \ref{mumediumTh}a one can see that the decoherence function
\eqref{mudeco} follows the function $\mu_{\tau_d<\tau_s}(t)$ for short times $t<\tau_s$ and for large times
$t\gg\tau_s$ oscillates around the approximation $\mu_{\tau_d>\tau_s}(t)$. The oscillations result from the rotation and {\sl breathing} of the Wigner function in phase space (with frequency $\omega_0\sim\tau_s^{-1}$) which is connected to a periodical change between superposition in coordinate and momentum space.
For low temperatures $kT\ll\hbar\omega_0$ the functions
\eqref{muGR} and \eqref{muWW} become
\begin{equation}
\mu_{\tau_d>\tau_s}(t)=e^{-(2\gamma
\omega_0q_0^2/\hbar)t}\qquad\mbox{and}\quad\mu_{\tau_d<\tau_s}(t)\sim\left\{\begin{array}{c}
e^{-\frac{8\gamma\Gamma^2q_0^2}{\pi\hbar}t^2}\quad\mbox{for}\quad t\ll\tau_s,\\
t^{-\frac{8\gamma q_0^2}{\hbar\pi}}\quad\mbox{for}\quad
t\gg\tau_s.\end{array}\right.\end{equation}
Figure \ref{mumediumTh}b shows, that the decoherence function
\eqref{mudeco} for short times $t<\tau_s$ decay fast in accordance with the evolution of
$\mu_{\tau_d<\tau_s}(t)$, while the long-time behavior for $t\gg\tau_s$ is again well approximated by $\mu_{\tau_d>\tau_s}(t)$. The long-time behavior of $\mu_{\tau_d<\tau_s}(t)$ however can be characterized by a power law decay $\sim t^{-\gamma q_0^2}$. This result is in accordance with the behavior of a free quantum Brownian particle \cite{sinha}. There is no characteristic decoherence time any more. In our case the decoherence function follows an algebraic decay for intermediate times and an exponential decay for large times.

\subsection{Numerical Analysis of the dynamics}
\subsubsection{Decoherence of a single-mode superposition}\quad\\
Describing decoherence processes within the Born-Markov approximation is valid as long as the time scale of bath correlations $\tau_b$ is the smallest time scale. Since the decoherence time scale $\tau_d$ is indirect proportional to the bath temperature $T$ and to the squared separation $q_0^2$, the decoherence process takes place on time scales that are comparable to the bath correlation time. In this case, non-Markovian influences become important. \\
\begin{figure}[t]
\begin{center}
\psfrag{t}[c]{$\omega_0t$}\psfrag{x}[c]{$\mu_I(t)$}
\subfigure[$\tau_d\gg\tau_b$ for $\Gamma=100\omega_0$,
$|\alpha_0|=100$.]{\includegraphics[width=5cm,clip]{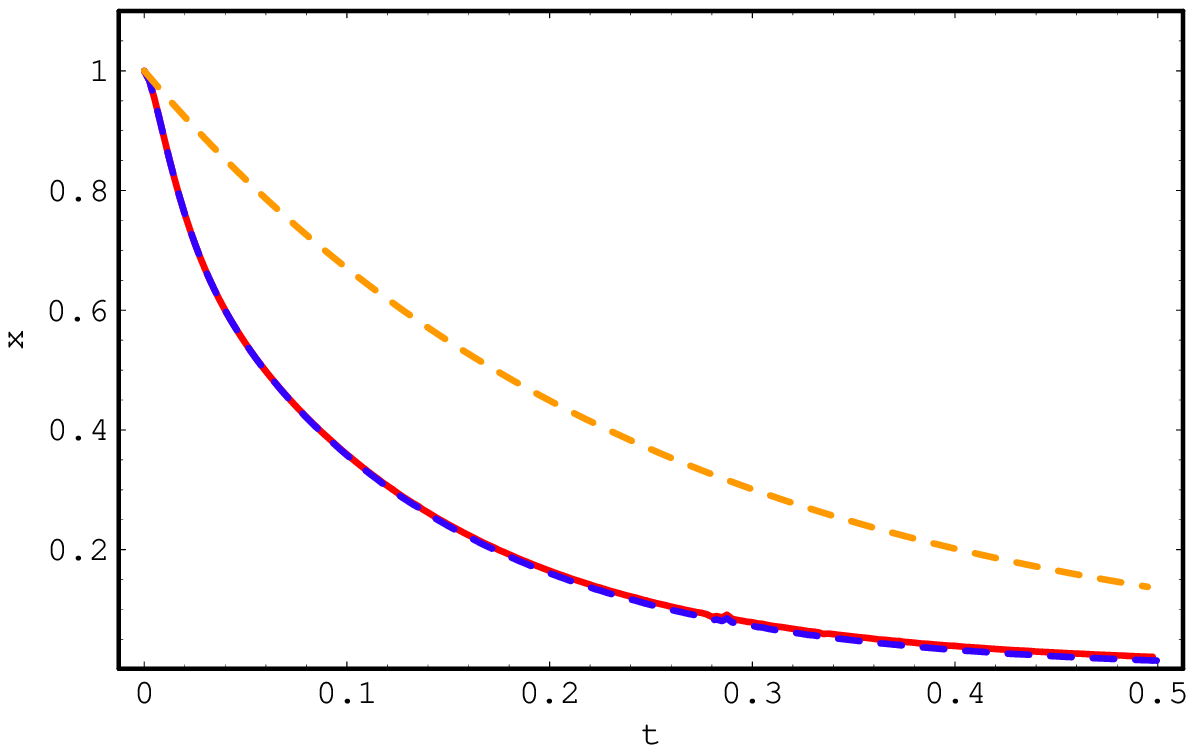}}\qquad
\subfigure[$\tau_d\approx\tau_b$ for $\Gamma=10\omega_0$,
$|\alpha_0|=100$.]{\includegraphics[width=5cm,clip]{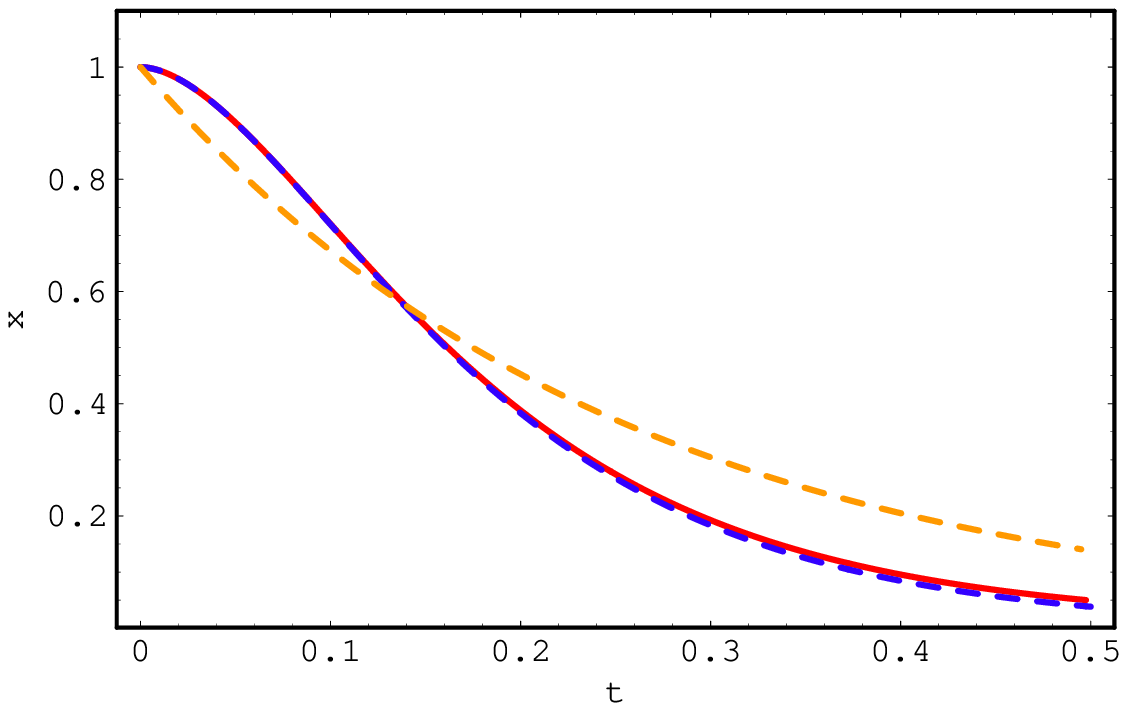}}\qquad\\
\subfigure[$\tau_d\approx\tau_b$ for $\Gamma=2\omega_0$,
$|\alpha_0|=100$.]{\includegraphics[width=5cm,clip]{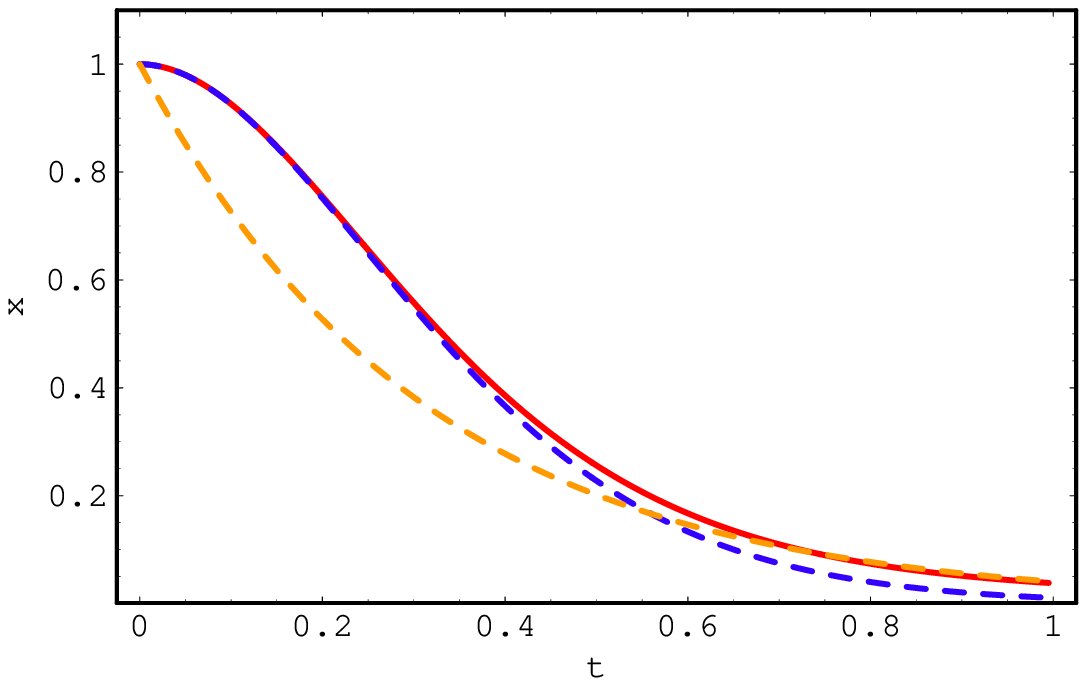}}\qquad
\subfigure[$\tau_d<\tau_b$ for $\Gamma=10\omega_0$,
$|\alpha_0|=500$.]{\includegraphics[width=5cm,clip]{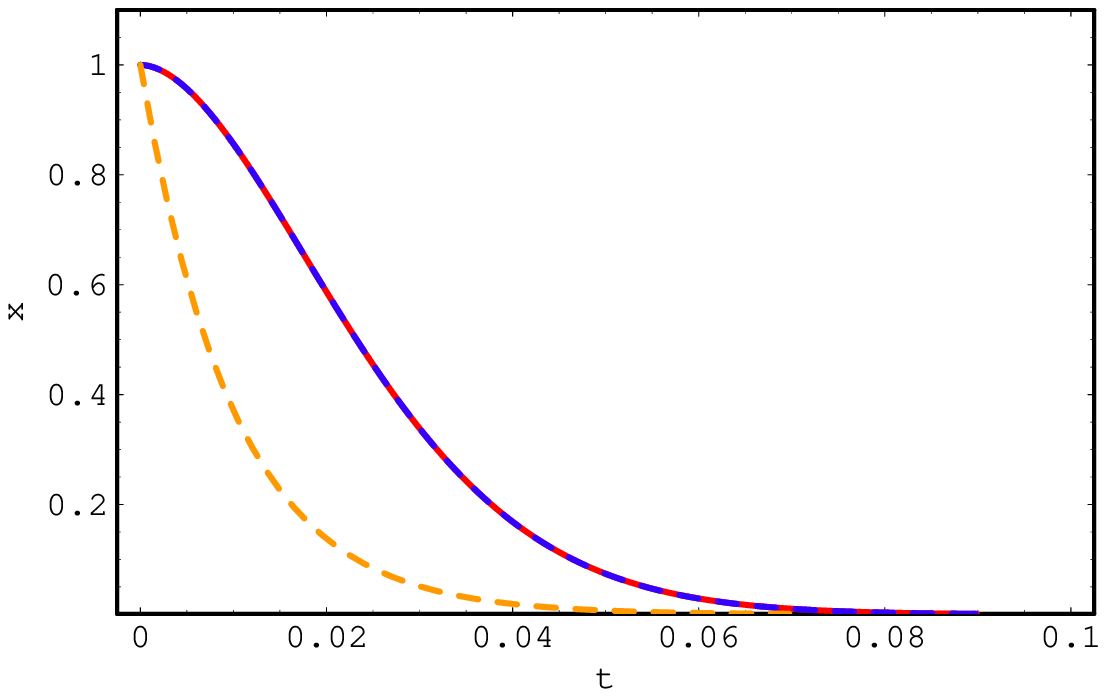}}\qquad\\
\subfigure[$\tau_b\ll\tau_s\ll\tau_d$ for $\Gamma=10\omega_0$,
$|\alpha_0|=10$.]{\includegraphics[width=5cm,clip]{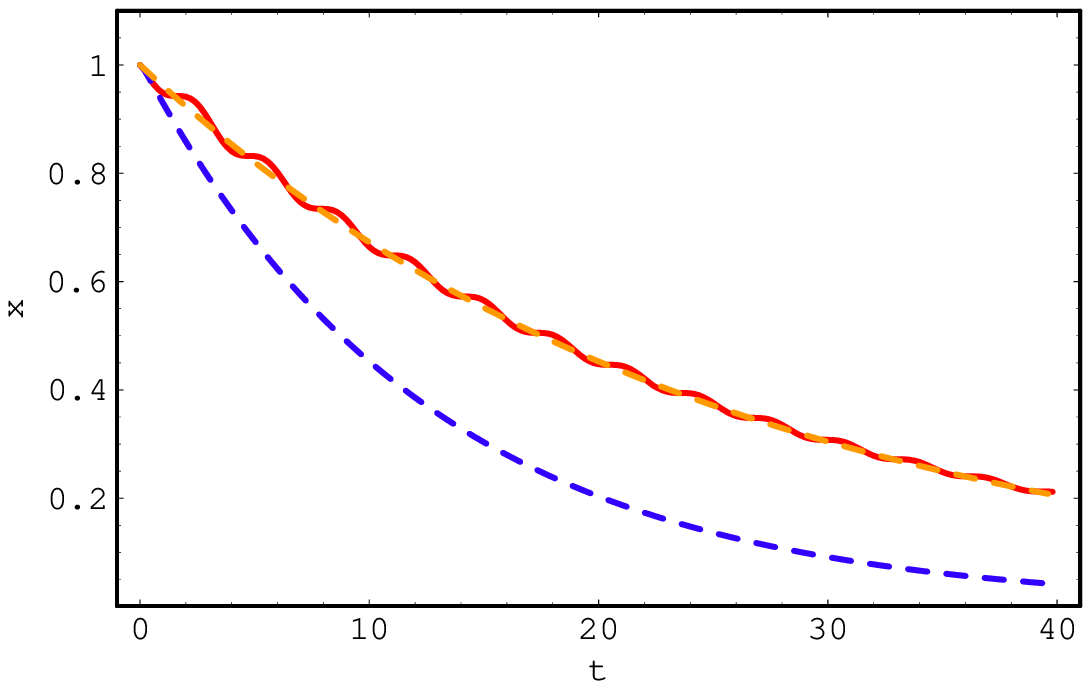}}\qquad
\subfigure[$\tau_b\gg\tau_s$ for $\Gamma=0.01\omega_0$, $|\alpha_0|=30$ and $\gamma=0.05\omega_0$ compared to $\gamma=0.1\omega_0$ (dashed lines).  ]{\includegraphics[width=5cm,clip]{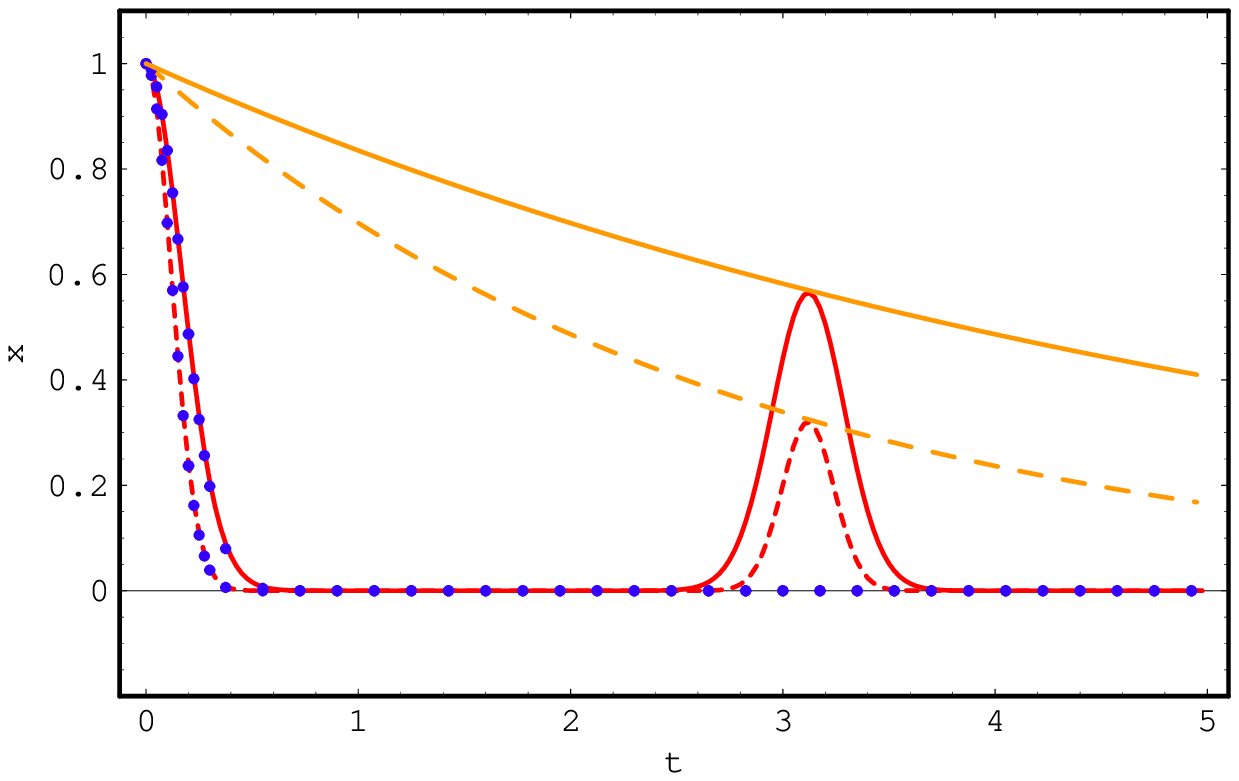}}
 \caption[]
 {\label{decononmarkov}\sl\small  Time evolution of the decoherence function \eqref{mudeco} (red lines) in comparison to the approximation \eqref{muWW} (blue lines) and \eqref{muGR} (yellow lines) at high temperature $kT=10\hbar\omega_0$ and small coupling $\gamma=10^{-5}\omega_0$ in $(a)$ to $(e)$. Figures $(a)$ to $(d)$
 illustrate the transition from the Markovian ($\tau_d\gg\tau_b$) to the non-Markovian regime
 ($\tau_d\ll\tau_b$), related to the ratio of decoherence time
 $\tau_d$ and bath correlation time scale $\tau_b\approx\Gamma^{-1}$. The decoherence function $\mu_I(t)$ follows the approximated function \eqref{muWW} since $\tau_d\ll\tau_s\sim\omega_0^{-1}$. A comparison with the case
 $\tau_d\gg\tau_s$, where $\mu_I(t)$ follows the function \eqref{muGR} is plotted in figure $(e)$.
 Figure $(f)$ illustrates the partial revivals of coherence in an out-of-resonance-bath.}
 \end{center}
\end{figure}

A gradual change from the Markovian to the non-Markovian regime can be studied by varying the bath correlation time in form of the inverse cutoff frequency  $\Gamma^{-1}$ and the decoherence time scale $\tau_d$ by the parameter $|\alpha_0|=q_0/2\sigma_0$. Such a change from the Markovian ($\tau_d\gg\tau_b$) to the non-Markovian regime
($\tau_d\ll\tau_b$) is illustrated in figures \ref{decononmarkov}a-f by the evolution of $\mu_I(t)$,
$\mu_{\tau_d>\tau_s}(t)$ and $\mu_{\tau_d<\tau_s}(t)$. Figures
\ref{decononmarkov}a-c show that a reduction of $\Gamma$ slows down the decoherence process. For example, the decoherence time in fig. \ref{decononmarkov}c for $\Gamma=2\omega_0$ is five times larger than the decoherence time in fig. \ref{decononmarkov}a for $\Gamma=100\omega_0$. More important are the qualitative changes in the evolution of the three decoherence measures $\mu_I(t)$, $\mu_{\tau_d>\tau_s}(t)$ and $\mu_{\tau_d<\tau_s}(t)$. In figures \ref{decononmarkov}a-d the norm $\mu_I(t)$ follows
$\mu_{\tau_d<\tau_s}(t)$, since the decoherence time is smaller than the characteristic system time scale $\tau_s\sim\omega_0^{-1}$. A comparison to the case $\tau_d\gg\tau_s$ is shown in figure \ref{decononmarkov}d where
$\mu_I(t)$ follows the evolution of $\mu_{\tau_d>\tau_s}(t)$. \\

So far, we have considered non-Markovian influences by approaching $\tau_d$ and $\tau_b$, where the limiting cases for $\tau_d\gg\tau_s$ and $\tau_d\ll\tau_s$ have been distinguished. A further distinctive feature can be observed if additionally the condition $\tau_b\gg\tau_s$ is fulfilled which means $\omega_0\gg\Gamma$. In this case the bath could be characterized as {\sl out-of-resonance} \cite{manisc1}. The relaxation timescale $\tau_{\gamma}\sim\gamma\omega_0^2/\Gamma^2$ also depends on the cut-off frequency but still remains by far the largest time scale. The time evolution of the decoherence function $\mu_I(t)$ cannot be well approximated by $\mu_{\tau_d<\tau_s}$ or $\mu_{\tau_d>\tau_s}$ alone. Figure \ref{decononmarkov}f shows an example. For short times the norm $\mu_I(t)$ decays very fast, following the approximated function $\mu_{\tau_d<\tau_s}$. However, within the half of a system period  $\omega_0t\approx\pi$ a partial revival of coherence takes place and the function $\mu_I(t)$ reaches a relative maximum that is given by the corresponding value of limit case $\mu_{\tau_d>\tau_s}$. Although the characteristic time scales of $\mu_{\tau_d<\tau_s}$ and $\mu_{\tau_d>\tau_s}$ are quite different, only both limit cases taken together give a accurate picture of the decoherence process in this regime. The corresponding master equation \eqref{maniseq} in this case is not of Lindblad-type with partially negative values of the coefficients $\tilde \gamma_{\uparrow}(t)$ and $\tilde \gamma_{\downarrow}(t)$.
\begin{figure}[h]
\begin{center}
\psfrag{t}[c]{$\omega_0 t$} \psfrag{x}[c]{$\mathcal{C}_{12}(t)$}
\subfigure[$\Gamma=3\omega_0$, $|\alpha_0|=20$.]{\includegraphics[width=5cm,clip]{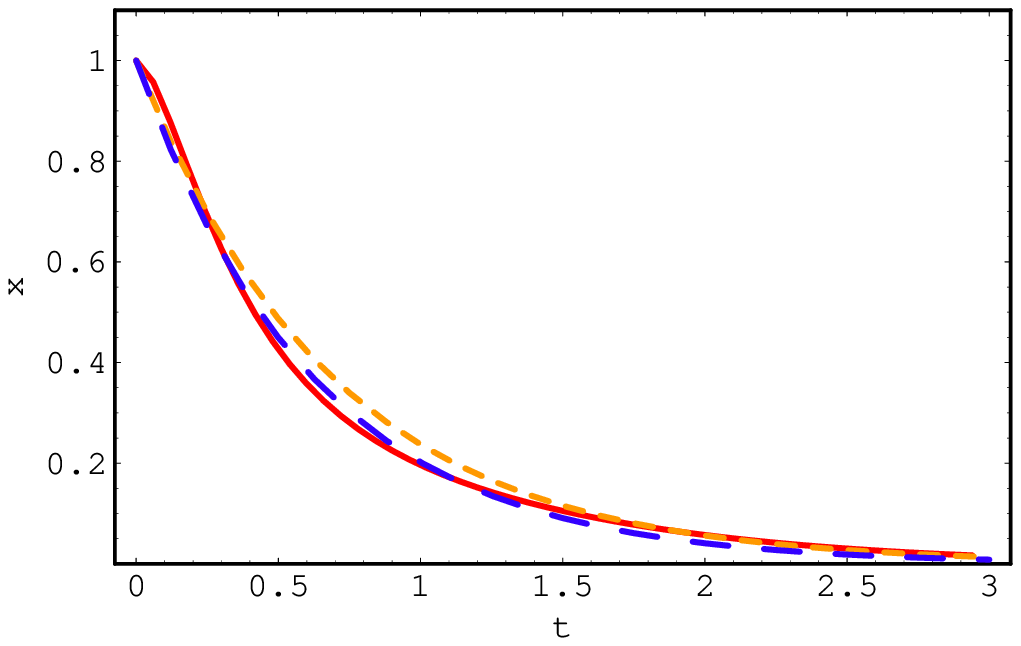}}\qquad
\subfigure[$\Gamma=0.5\omega_0$, $|\alpha_0|=20$.]{\includegraphics[width=5cm,clip]{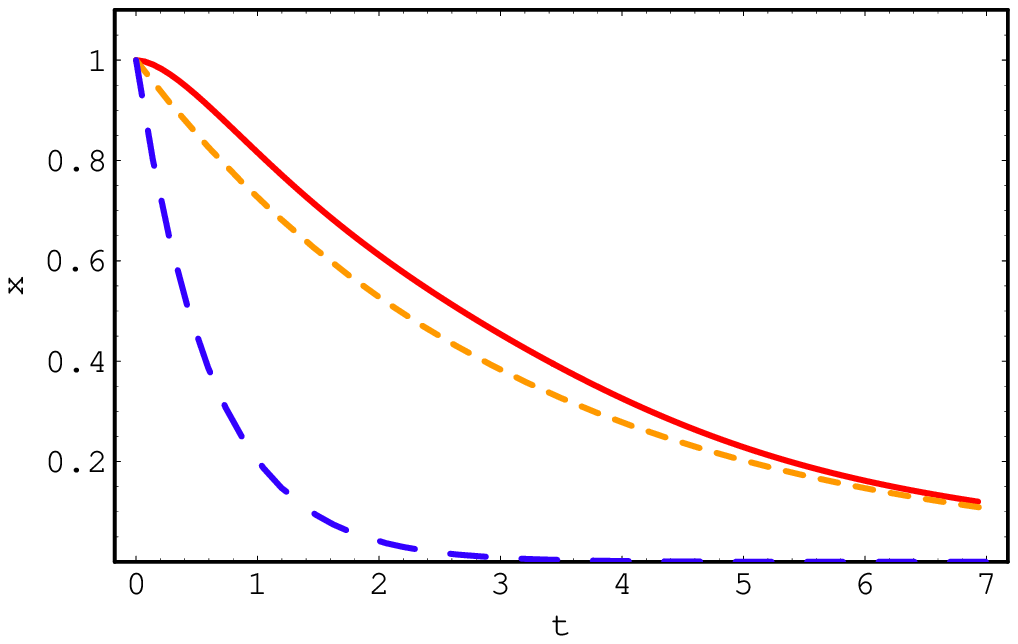}}\qquad\\
\subfigure[$\Gamma=10^{-2}\omega_0$, $|\alpha_0|=1500$.]{\includegraphics[width=5cm,clip]{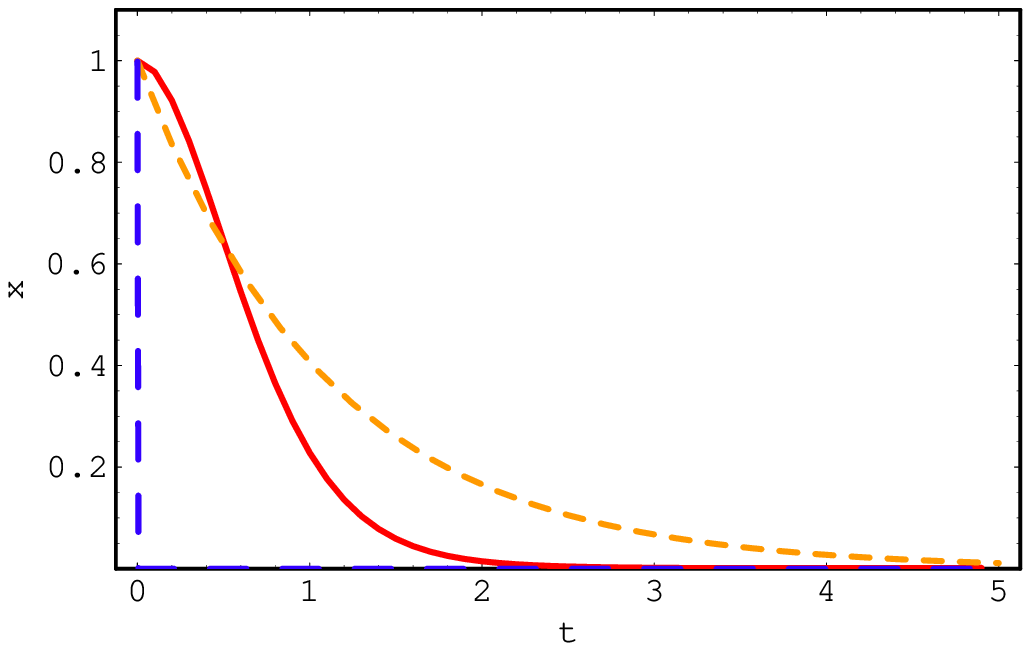}}\qquad
\subfigure[$\Gamma=10^{-3}\omega_0$, $|\alpha_0|=3500$.]{\includegraphics[width=5cm,clip]{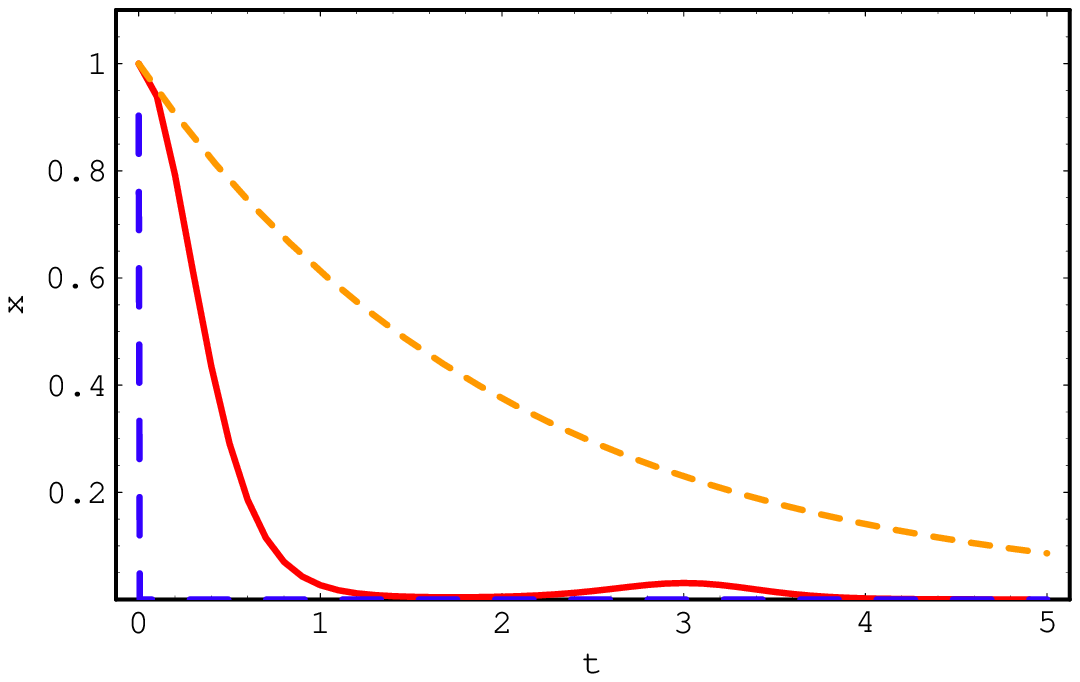}}
 \caption[]{\label{concurnm}\sl \small Time evolution of the concurrence $C_{12}(t)$ (red line) for a two-mode entangled coherent state $|\Phi_+\rangle$ \eqref{phialphacat} in a zero temperature environment with small ratios $\Gamma/\omega_0$ and $\gamma=10^{-3}\omega_0$.  In comparison the Markovian evolution is plotted for relaxation time $\gamma^{-1}$ (blue line) and with adjusted relaxation time $\tau_{\gamma}$ (yellow dashed line). In a moderate non-Markovian regime the entanglement is preserved for longer times $(b)$ while it is lost faster in an out-of-resonance bath with $\Gamma\ll\omega_0$ (c). The behavior of the concurrence is similar to that of the decoherence function. Revivals can occur even at zero temperature (d). }
 \end{center}
\end{figure}

\subsubsection{Disentanglement of a two-mode entangled coherent states}\quad\\
In the following, we compare the findings on the decoherence process of a single-mode superposition state to the behavior of the concurrence of a two-mode entangled coherent state. Figures \ref{concurnm}a-d show the time evolution of the concurrence $C_{12}(t)$ of an entangled coherent state $|\Phi_+\rangle=|\alpha,\theta=0,N=2\rangle$ in reservoirs with different specification of their non-Markovian character in form of the relation $\Gamma/\omega_0$. Qualitatively the time evolution resembles strongly the results for a single-mode superposition cat state. Starting from a quasi Ohmic bath with ($\Gamma\gg\omega_0$) in fig. \ref{concurnm}a) a reduction of the cut off frequency leads to deviations between the Markovian and non-Markovian results. For moderate non-Markovian influences the entanglement is preserved most efficiently (fig.\ref{concurnm}b). In the strongly non-Markovian out-of-resonance bath with $\Gamma<\omega_0$ this relation is only valid at short times while for longer times the non-Markovian concurrence even decays faster, as can be seen from fig. \ref{concurnm}c. The occurrence of coherence revivals is similar to the results found for the decoherence function of a single-mode superposition of coherent states.

\section{Summary and Conclusions}
In this paper we have analyzed the non-Markovian effects on decoherence and disentanglement processes of non-Gaussian continuous variable systems within the quantum Brownian motion model. We have compared the time evolution of the decoherence function of a single-mode cat state with the evolution of the concurrence of a two-mode entangled coherent state. For both cases we studied different decoherence and disentanglement scenarios depending on the relation between the characteristic time scales of system and environment. The entanglement dynamics of two-mode entangled coherent states is similar to the decoherence dynamics of single-mode coherent cat-states. We found exponential, Gaussian and algebraic decay patterns of the decoherence function in moderate non-Markovian regime and revivals of decoherence and concurrence in strongly non-Markovian out-of-resonance reservoirs.

\end{document}